\documentclass[aps,prl,twocolumn,showpacs,subfigure,superscriptaddress,nobibnotes,nofootinbib]{revtex4-1}
\usepackage[dvips]{graphicx}
\usepackage{epsfig}
\usepackage{bm}   
\usepackage{dcolumn}
\usepackage{graphicx}
\usepackage{rotate}
\usepackage{amsfonts}
\usepackage[table]{xcolor}
\usepackage{tabularx}
\usepackage{footmisc,booktabs,amssymb}

\definecolor{capri}{rgb}{0.0, 0.75, 1.0}
\definecolor{cornflowerblue}{rgb}{0.39, 0.58, 0.93}
\definecolor{spirodiscoball}{rgb}{0.06, 0.75, 0.99}
\definecolor{pear}{rgb}{0.82, 0.89, 0.19}

\begin{document}

\title{Competition between $(\gamma,p)$ and $(\gamma,n)$ photo-disintegration yields}

\author{Jos\'e Nicol\'as Orce}
\email{jnorce@uwc.ac.za}
\homepage{http://nuclear.uwc.ac.za/} 
\affiliation{Department of Physics \& Astronomy, University of the Western Cape, P/B X17, Bellville, ZA-7535 South Africa}
\affiliation{National Institute for Theoretical and Computational Sciences (NITheCS), South Africa}

\date{\today}

\begin{abstract}

A comprehensive analysis of the competition between photo-proton and photo-neutron disintegration yields is undertaken by constraining 
photo-absorption data with physics principles -- namely, dipole sum rules, decay properties to open channels, isospin selection rules 
and statistical evaporation from compound nucleus formation. 
The trend and magnitude of disintegration yields for self-conjugate nuclei are in agreement with 
the evaporation model of Blatt and Weisskopf. A substantial improvement is found between the exponential trend presented in 
this work and that determined by Morinaga using a similar approach. 
This work allows for the evaluation of frequently missing or inconsistent photo-proton contributions, 
which  impacts the photo-disintegration of light and neutron-deficient nuclei, where photo-proton 
contributions are relevant.

\end{abstract}



\maketitle

\section{Motivation}

\begin{figure}
\begin{center}
\includegraphics[width=8cm,height=6cm,angle=-0]{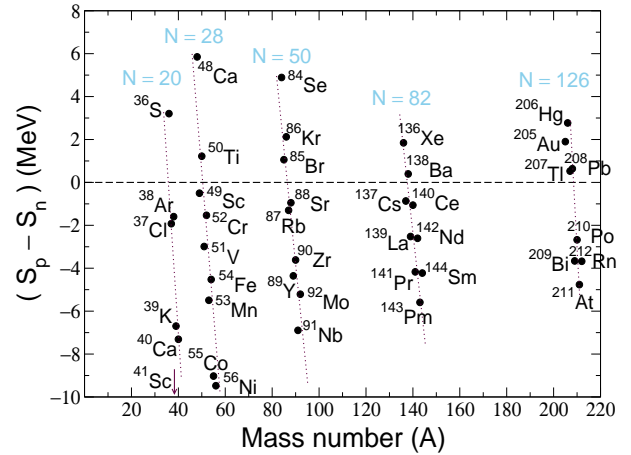} 
\caption{(Color online) Separation energy difference between protons and neutrons $(S_p - S_n)$ as a function of mass number $A=N+Z$ for 
the $N = 20$, 28, 50 82 and 126 isotones (black circles). 
Mass data are taken from the 2020 atomic mass evaluation~\cite{mass2020}.} 
\label{fig:SpminusSn}
\end{center}
\end{figure}

Photo-disintegration occurs when a nucleus is heated by the absorption of a photon and decays into another nucleus 
by the evaporation of one or several particles~\cite{photodisintegration}. 
Because of the high Coulomb barrier, neutron emission is generally the predominant decay mode in heavy nuclei. 
Competing proton emission can be relevant for light and neutron-deficient nuclei; particularly those 
with neutron magic numbers where, as shown in Fig.~\ref{fig:SpminusSn},  proton separation energies ($S_p$) lie much lower than neutron thresholds  ($S_n$). 

Such a competition between photo-proton $\sigma(\gamma,p)$ and photo-neutron $\sigma(\gamma,n)$ 
cross sections impacts a variety of nuclear and astrophysics phenomena --- namely, bremsstrahlung~\cite{migdal,levinger}, nuclear polarizability~\cite{orce_review}, three-nucleon forces~\cite{hebeler},
collective properties~\cite{eichler}, isospin asymmetry and symmetry energy in neutron stars~\cite{neutronstars,latimer,latimer2,pearson}, 
neon and silicon burning at the end of stellar evolution~\cite{stellarmodels}  and the 
nucleosynthesis of $p$-nuclei~\cite{pnuclei1,pnuclei2}, which are shielded from the rapid-neutron 
capture by stable isobars. A recent review article by Zilges and co-workers discusses in detail various applications 
of photonuclear physics~\cite{photonuclear_review}, e.g. finding new routes for production of medical radioisotopes or 
nuclear-waste transmutation. 

Most of the absorption or emission of $\gamma$ rays in nuclei arises from the giant dipole resonance ({\small GDR})~\cite{atlas,Kawano}, 
originally predicted as interpenetrating proton and neutron fluids moving collectively out of phase~\cite{migdal,migdal75},
and soon after discovered 
as the first quantum collective-excitation mode in nuclear physics~\cite{GDR_discovery}. 
 The understanding of the {GDR} follows a fascinating path~\cite{danos} from its original hydrodynamic interpretation as a
liquid drop~\cite{migdal} to its complementary shell-model representation as a system of
independent nucleons plus a residual interaction~\cite{levinger3,balashov,danosfuller}. 

\begin{table*}[!ht]
\caption{Experimental photo-proton and photo-neutron cross sections\footnote{Typical 
uncertainties of 15\% for $\sigma(\gamma,p)$ and 10\% for  $\sigma(\gamma,n)$ cross sections are considered. \\
Units are in MeV$\cdot$mb 
for integrated cross sections and MeV for energies and particle thresholds.} with the corresponding $r_{pn}^{exp}$ experimental ratio, total empirical and {\small TRK} cross sections, several variables explained in the text and shown in Fig.~\ref{fig:gpgn} and $r_{pn}$ values extracted from the evaporation model, open channels, $\frac{N_p}{N_n}$, and isospin intensities, $\frac{I(T+1)}{I(T)}$.}
\label{tab:rpntab}
\begingroup\setlength{\fboxsep}{0pt}
\colorbox{gray!70!yellow!10}{%
\begin{tabular}{|c|c|c|c|c|c|c|c|c|c|c|c|}
\hline 
  Nucleus    &   Ref.                          &  $\sigma(\gamma,p)$  &  $\sigma(\gamma,n)$ &  $r_{pn}^{exp}$  & $\sigma_{_{0}}$ &  $\sigma^{TRK}_{_{0}}$ & $e^{a((S_n+\varepsilon_n)-(S_p+\varepsilon_p))}$  &  $S_p-S_n$  &  $E_{X_p}-E_{X_n}$  &  $\frac{N_p}{N_n}$  & $\frac{I(T+1)}{I(T)}$  \\ \hline
 $^{12}$C    & \cite{fuller,hanna,lightnuclei} & 110  &   54  &  2.0(4)   &  164(17)  &  180  &  1.1 &  -2.76 & 0.619  &   1.0  &  2.3  \\  
 $^{16}$O    & \cite{fuller,hanna,lightnuclei} & 133  &   81  &  1.7(3)   &  214(22)  &  240  &  1.1 &  -3.54 & 0.727  &   1.0  &  2.4  \\  
 $^{20}$Ne   & \cite{gorbunov}                 & 231  &  115  &  2.0(4)   &  346(37)  &  300  &  1.3 &  -4.02 & 0.601  &   1.0  &  2.5  \\  
 $^{24}$Mg   & \cite{varlamov24mg,lightnuclei} & 180  &   76  &  2.4(6)   &  310(50)  &  360  &  1.8 &  -4.84 & 0.846  &   1.8  &  2.5  \\  
 $^{28}$Si   & \cite{lightnuclei}              & 230  &   90  &  2.6(4)   &  350(40)  &  420  &  2.5 &  -5.59 & 1.062  &   1.7  &  2.6  \\   
 $^{32}$S    & \cite{lightnuclei}              & 350  &  112  &  3.1(6)   &  410(40)  &  480  &  3.1 &  -6.18 & 1.133  &   2.1  &  2.6  \\  
 $^{36}$Ar   & \cite{lightnuclei}              &      &       &           &           &  540  &  3.7 &  -6.75 & 1.207  &   3.0  &  2.6  \\  
 $^{40}$Ca   & \cite{lightnuclei}              & 470  &   88  &  5.3(7)   &  560(90)  &  600  &  4.6 &  -7.31 & 1.289  &   3.5  &  2.7  \\ \hline
\end{tabular}
}\endgroup
 \end{table*}

Previous photo-absorption data have been under question since 1988, when Dietrich and Berman evaluated the $\sigma(\gamma,n)$ 
data~\cite{atlas} from measurements 
using monochromatic photon beams
generated by in-flight annihilation of positrons~\cite{lectures}. 
Systematic discrepancies between data obtained by Bremsstrahlung and quasi-monoenergetic beams arise mainly at energies above the {\small GDR}
region~\cite{iskhanov,gheorghe} because of the complicated process of 
sorting neutrons in multiplicity, i.e. how the $(\gamma,n)$, $(\gamma,2n)$, $(\gamma,3n)$,... partial cross sections are
unfolded~\cite{varlamov1}. 
The solution arises once systematic uncertainties are properly
considered by decomposing experimental total neutron yield reaction
cross sections into partial reaction contributions~\cite{gheorghe,varlamov2,varlamov3}. 
A collaborative project led by the International Atomic
Energy Agency (IAEA) has recently commissioned an important update to the photo-absorption  data
library~\cite{Kawano,iaea}.

The situation is far more unsettled when $\sigma(\gamma,p)$ contributions are concerned, 
where $\sigma(\gamma,p)$ data are either scarce or highly inconsistent. See e.g. Ref.~\cite{allen} and references therein, where 
the absolute $\sigma(\gamma,p)$ in $^{20}$Ne ranges from 61(11) to 165(25) MeV $\cdot$ mb. An even smaller cross section is determined if 
the emitting proton is assumed to leave the residual nucleus in its ground state, $\sigma(\gamma,p)=28(9)$ MeV $\cdot$ mb~\cite{gp28,gp25}. 
Although it is evident that much more accurate $\sigma(\gamma,p)$ data are needed with current state-of-the-art facilities such as 
HIGS~\cite{higs}, ELBE~\cite{elbe} and the modern ELI-NP~\cite{eli},  
further means to estimate the most likely $\sigma(\gamma,p)$  contributions are required. 
This is precisely the purpose of this work.\\

\section{Testing ground}

A comprehensive testing ground is provided by $A=4n$  self-conjugate nuclei, i.e. nuclei with the same number of 
protons and neutrons, where proton emission can be the dominant decay mode for two main reasons: 
 \begin{enumerate}
  \item Lower proton than neutron thresholds;  hence, a greater number of available open proton channels. 
  \item Isospin selection rules favor proton decay -- as explained below -- from the decay of the isospin $T=1$ {\small GDR} resonance~\cite{lectures,orce1,morinaga2,mahaux_sdshell,shoda}.  
   \end{enumerate}

Let's now define  the photo-proton to photo-neutron yield ratio  $r_{pn}$ as  
 \begin{equation}
   r_{pn}= \frac{\sigma(\gamma,p)}{\sigma(\gamma,n)},
 \end{equation}
where both $\sigma(\gamma,p)$ and $\sigma(\gamma,n)$ include all the partial channels 
and  satisfy the dipole (or Thomas–Reiche–Kuhn ({\small TRK})) sum rule for the total 
photo-absorption cross section~\cite{levingerbethe,levinger2}, 
\begin{eqnarray}
\sigma_{_0} & = & \int_0^\infty{\sigma dE} = \sigma(\gamma,p) + \sigma(\gamma,n)  \nonumber \\
& = &\frac{2\pi^2e^2\hbar c}{Mc^2} \frac{NZ}{A} (1+\Delta) = 60 \frac{NZ}{A} (1+\Delta), 
\end{eqnarray}
where $\Delta$ is the contribution of exchange forces above the pion threshold at 140 MeV and calculated to be in the range $0.4-1.0$~\cite{levingerbethe,rand,johnson}. Henceforth,  additional contributions such as $\sigma(\gamma,\alpha)$  (e.g. $\approx1\%$ for $^{20}$Ne~\cite{gorbunov}) are assumed negligible; although these may have a larger contribution when considering other sum rules such as 
$\sigma_{_{-2}} = \int{\sigma E^{-2} dE}$, where the $E^{-2}$ factor has to be considered~\cite{20ne_mehl}. 
Although out of the scope of this work, the $\frac{\sigma(\gamma,p)}{\sigma(\gamma,n)}$ ratio can also provide information on the role of the different photo-disintegration mechanisms -- namely, semidirect and statistical decays~\cite{24Mg,lightnuclei}. The former is related to collective 1-particle-1-hole ($1p1h$) excitations, whereas the latter includes all pre-equilibrium and equilibrium processes and involves all possible $np-nh$ excitations.

\begin{figure}[!ht]
\begin{center}
\includegraphics[width=7.8cm,height=6.cm,angle=-0]{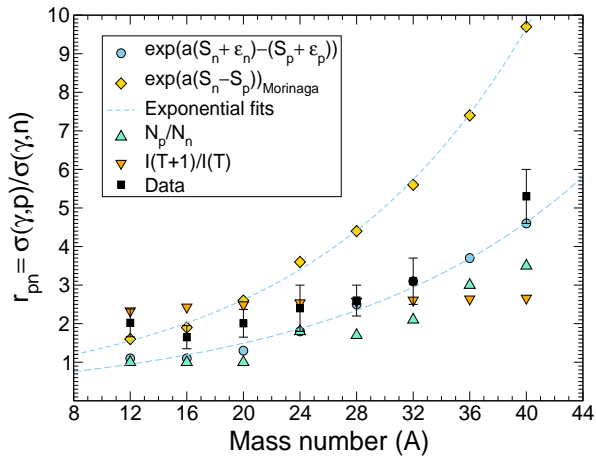} 
\caption{(Color online) Experimental $r_{pn}$ yield ratios (black squares) for particular $A=4n$ self-conjugate nuclei compared with $r_{pn}$ values determined using different approaches -- namely, evaporation model (exponential trends), open channels ($\frac{N_p}{N_n}$) and isospin selection rules ($I(T+1)/I(T)$).} 
\label{fig:rpn}
\end{center}
\end{figure}

Table~\ref{tab:rpntab} and Fig. \ref{fig:rpn} show the empirical $r_{pn}$ values (squares) for $A=4n$ self-conjugate nuclei 
from $^{12}$C to $^{40}$Ca.  The consistency of data is crucial for this analysis and generally involve the selection of 
at least two experiments and an upper limit of integration of the cross sections set to 30 MeV~\cite{lightnuclei,exfor};  
except for $^{20}$Ne, where $(\gamma,p)$ and $(\gamma,n)$ contributions were evaluated from the same experiment~\cite{gorbunov}. 
Despite dipole absorption extends well beyond 30 MeV~\cite{gorbunov,komar,gorbunov2,woodworth},  Table~\ref{tab:rpntab} illustrates that 
the total photo-absorption cross-sections $\sigma_{_{0}}$  exhaust, within uncertainties, the 
{\small TRK} sum rule in most of the cases. 
If $\sigma(\gamma,p)$ is unavailable or need further testing, one could also determine it by subtracting the well-known $\sigma(\gamma,n)$ contribution from the total photo-absorption cross section~\cite{ahrens}. The latter, however, is even more scarcely available than $\sigma(\gamma,p)$ 
and were not utilized.  No data are available for $^{36}$Ar.

Furthermore, the consistency of the selected data can be benchmarked by physics principles --- namely,  
decay properties to open channels, isospin selection rules and statistical evaporation from compound nucleus formation.

\begin{figure}[!ht]
\begin{center}
\includegraphics[width=8.5cm,height=9.5cm,angle=-0]{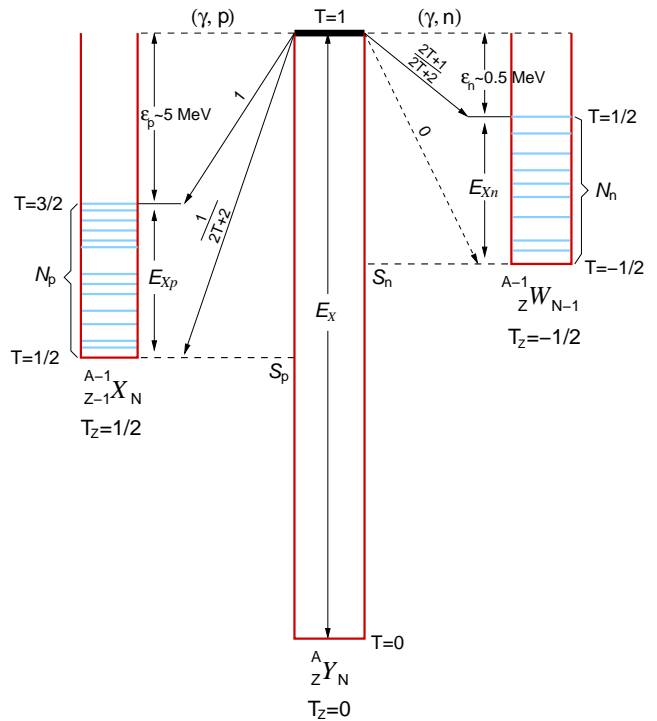} 
\caption{(Color online) Schematic diagram showing photo-absorption and photo-disintegration processes.} 
\label{fig:gpgn}
\end{center}
\end{figure}

\section{Open channels}

Figure~\ref{fig:gpgn} shows the schematic diagram for the 
photo-disintegration of a nucleus $^{A}_{Z}Y_N$. 
In principle, the $\frac{\sigma(\gamma,p)}{\sigma(\gamma,n)}$ ratio may be controlled by the magnitude of $S_p$ and $S_n$, and the number of open channels or levels, $N_p$ and $N_n$, in the residual nuclei $^{A-1}_{Z-1}X_N$ and $^{A-1}_{\ ~~Z}W_{N-1}$ below the excitation energies, $E_{X_p}$ and $E_{X_n}$,  respectively, such as~\cite{lectures}, 
 \begin{equation}
   r_{pn} = \frac{N_p}{N_n}. 
 \end{equation}
The number of discrete levels to take into account lie well below the quasi-continuum, with maximum excitation energies given by
\begin{eqnarray}
E_{X_p} &=& E_{_{X}} - S_p - \varepsilon_p, \\ 
E_{X_n} &=& E_{_{X}} - S_n - \varepsilon_n,
\end{eqnarray} 
where $E_{_{X}}\approx E{_{GDR}}$, the kinetic energy of the protons is assumed to be the minimum kinetic energy to overcome the Coulomb barrier, 
$\varepsilon_p=\frac{Ze^2}{1.2A^{1/3}}$ and for the neutrons, a minimum kinetic energy of $\varepsilon_n=0.5$ MeV is 
considered.

The difference between $S_p$ and $S_n$  has been identified as the main factor responsible for the difference between $\sigma(\gamma,p)$ and 
$\sigma(\gamma,n)$ yields~\cite{24Mg,lectures}.
For instance, it is well known that $^{58}$Ni prefers to decay by
proton emission to $^{57}$Co, with a higher level density, than by neutron emission to the semi-magic nucleus $^{57}$Ni~\cite{lectures}. 
The opposite occurs for semimagic nuclei with a neutron magic number, where the $(\gamma,n)$ decay is favorable. 
The $\frac{N_p}{N_n}$ ratio has been found to follow the $\frac{\sigma(\gamma,p)}{\sigma(\gamma,n)}$ trend for $N=50$ semi-magic nuclei~\cite{shoda_92Mo}, $sd$-shell nuclei~\cite{lectures} and others (see e.g. the special 
cases of $^{58}$Ni and $^{92}$Mo~\cite{lectures}). 

The $\frac{N_p}{N_n}$ ratios are  listed in Table~\ref{tab:rpntab} and shown in Fig.~\ref{fig:rpn}. 
The results for $r_{pn} = \frac{N_p}{N_n}$ vary smoothly from 1 in $^{12}$C and $^{16}$O to 3.5 in $^{40}$Ca, in good agreement with the 
$\frac{\sigma(\gamma,p)}{\sigma(\gamma,n)}$ trend although not with the magnitude. In fact, it is known that 
the $\frac{N_p}{N_n}$ ratio is representative of $\frac{\sigma(\gamma,p)}{\sigma(\gamma,n)}$ for the lower energy half of the {\small GDR} and it is associated with the statistical nature of the levels in the residual nuclei~\cite{lectures}.

\section{Isospin selection rules}

Isospin selection rules remain valid in photonuclear reactions~\cite{murray}, where the {\small GDR} can 
generally split into isospin $T$ and $T+1$ components.  Nonetheless, $T=0 \rightarrow T=0$ transitions are not allowed 
in self-conjugate nuclei (with $T_{_Z} = \frac{N-Z}{2}=0$) by isospin selection rules~\cite{isospin} and only states with $T=1$ are excited by electric-dipole absorption.
If isospin symmetry is conserved, proton and neutron emission from the $T=1$ states of the {\small GDR} should be symmetric,  
with identical states in the residual mirror nuclei, i.e. $\frac{\sigma(\gamma,p)}{\sigma(\gamma,n)}=1$. 

Instead,  isospin symmetry is broken by the Coulomb force and the sum rules for transition strengths, shown in Fig.~\ref{fig:gpgn}, 
favors the  $(\gamma,p)$ decay  from the $T=1$ resonance~\cite{french,shoda}. 
This, together with the fact that neutron emission to the $T=1/2$ residual states is highly suppressed 
because of their expected high excitation energies  -- i.e. those transitions with the $\frac{2T+1}{2T+2}$ Clebsch-Gordan coefficient -- 
result in a favorable $(\gamma,p)$ contribution for $T_{_Z} =0$ nuclei.  
Evidence for a $T=0$ resonance decay has been used as a probe of isospin mixing in $T_{_Z} =0$ nuclei~\cite{hanna,mahaux_sdshell}, where the admixture of the $T=0$ to the $T=1$ states can be inferred from the 
$\frac{\sigma(\gamma,p)}{\sigma(\gamma,n)}$ ratio and found to be less than 4\%~\cite{lightnuclei}.  

In general, the relative intensities $I(T+1)$ and $I(T)$ for the $T+1$ and $T$ {\small GDR} decays can be associated with the  $r_{pn}$ ratio~\cite{fallieros}. Here, for the  particular case of $T_{_Z}=0$ nuclei, we propose the following equation, 
 \begin{eqnarray}
 r_{pn} &=&\frac{I(T+1)}{I(T)}=\frac{(1+ \frac{1}{2T+2})(2T+2)}{2T+1}\frac{1-\frac{3}{2}TA^{-2/3}}{1+\frac{3}{2}A^{-2/3}} \nonumber \\
        &=& \frac{3}{1+\frac{3}{2}A^{-2/3}},
 \end{eqnarray}
where, as shown in Fig.~\ref{fig:gpgn}, the weighting factors for  the $(\gamma, p)$ and $(\gamma, n)$ channels are 
given by the respective Clebsch-Gordan coefficients. 
Although the $\frac{I(T+1)}{I(T)}$ ratios increased smoothly from 2.3 in $^{12}$C to 2.7 in $^{40}$Ca, they do not follow the exponential 
data trend. Similar disagreements were encountered in previous work, where no weighting factors were considered~\cite{lectures}.

\section{Evaporation model}
 
The emission of protons and neutrons in photo-disintegration reactions can be described fairly well by the evaporation model~\cite{blattweisskopf}, 
where the nucleus is heated by the absorption of a photon at an excitation energy $E_x$ and  decays by the evaporation of one or several particles according to the statistical theory~\cite{sven,morinaga_evaporation}.

Assuming compound-nucleus formation and isospin symmetry --  and ignoring direct photoemission effects -- Morinaga found that the 
asymptotic value of the $r_{pn}$ ratio for  $A=4n$  self-conjugate nuclei  can be given by~\cite{morinaga_evaporation} 
\begin{equation}
r_{pn} = \frac{e^{a(E_x-(S_p+\varepsilon_p))}-1}{e^{a(E_x-(S_n+\varepsilon_n))}-1}\approx e^{a[(S_n+\varepsilon_n)-(S_p+\varepsilon_p)]}, 
\label{morinaga}
\end{equation}
where $a$ is the level density parameter~\cite{blattweisskopf}, $E_x$ the excitation energy of the compound nucleus, 
$S_p$ and $S_n$ the proton and neutron separation energies extracted from the 2020 atomic mass evaluation~\cite{ame2020},   $\varepsilon_p$ and $\varepsilon_p$ the kinetic energy of the 
outgoing proton and neutron, respectively, where $\varepsilon_n\approx0.5$ MeV and $\varepsilon_p$ is the minimum 
energy needed to overcome the Coulomb barrier, which ranges from 3-7 MeV between $^{12}$C and $^{40}$Ca. This difference 
in the kinematic energies have been found as the main factor for the
difference between the proton and neutron photo-disintegration yields in self-conjugate nuclei~\cite{lightnuclei}.

Although there are different  prescriptions for the level density parameter~\cite{a1,a3,a2}, 
this has preferably to be determined experimentally as it depends on the low-lying nuclear structure of the particular residual nucleus. 
Here, level density parameters have simply been determined by counting the number of levels within the excitation energies 
$E_{X_p}$ and $E_{X_n}$. 

The $r_{pn}$ ratio provided by $e^{a[(S_n+\varepsilon_n)-(S_p+\varepsilon_p)]}$ is plotted in Fig.~\ref{fig:rpn} (circles) together with an 
exponential fit (dashed line)  and listed in Table~\ref{tab:rpntab}. 
For comparison, Fig. \ref{fig:rpn} also shows the corresponding results (diamonds) and exponential fit 
to the asymptotic $r_{pn}$ values determined by Morinaga in 1955~\cite{morinaga2}, which shows increasingly anomalous $r_{pn}$ values 
as the mass number $A$ increases.

\begin{figure}[!ht]
\begin{center}
\includegraphics[width=7.cm,height=6.cm,angle=-0]{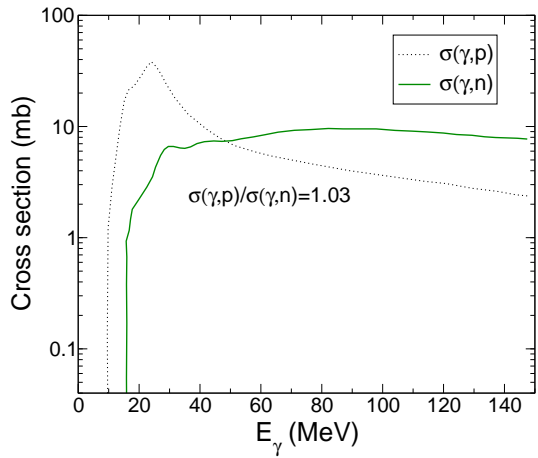} 
\caption{(Color online) Protoproton and photoneutron cross sections calculated with {\small GNASH} for the $\gamma~+$ $^{36}$Ar 
photodisintegration reaction. 
Figure modified from Ref.~\cite{gnash3}. Similar $r_{pn}\approx1$ ratios are calculated for $^{24}$Mg and $^{28}$Si.}
\label{fig:gnash}
\end{center}
\end{figure}

More sophisticated computer codes such as {\small HMS-ALICE}~\cite{hms}, {\small GNASH}~\cite{gnash1,gnash2}, {\small TALYS}~\cite{talys}, and {\small EMPIRE}~\cite{empire} can be used to estimate different photo-absorption cross sections. In particular, {\small GNASH} uses Hauser-Feshbach theory with multiple particle emission and includes corrections for pre-equilibrium effects. It ranks among the best codes 
to treat photonuclear reactions~\cite{bestcode} and has been used extensively to calculate neutron, proton, deuteron, triton and alpha production yields~\cite{gnash3} using the Korea Atomic Energy Research Institute ({\small KAERI}) library~\cite{kaeri}, which provides photonuclear data of 143 isotopes from $^{12}$C to $^{208}$Pb. Although it generally yields good results for total photo-absorption cross sections, it consistently underestimates  the $r_{pn}$ ratio for self-conjugate nuclei~\cite{gnash3}, as shown in Fig.~\ref{fig:gnash} for the photodisintegration of 
$^{36}$Ar, providing almost constant values of $\frac{\sigma(\gamma,p)}{\sigma(\gamma,n)}\approx1$ for $^{24}$Mg, $^{28}$Si and $^{36}$Ar, in disagreement with the trends provided by data and the simple evaporation-model assumptions presented in this work for $A=4n$ self-conjugate nuclei. Better results may be obtained using one of the three options for level densities available in {\small GNASH} or with different codes, but this kind of study is outside the scope of this work.

Finally, this work provides a means to determine the so-often missing or discrepant photo-proton yields by constraining data 
with physics principles and motivates further systematics studies in modern facilities such as {\small ELI-NP}. 
In particularly, it is relevant to explore the exponential trend predicted by the evaporation model with heavier 
neutron-deficient nuclei and where it evolves in favor of photo-neutron cross sections. Although there is some $(\gamma,p)$ 
information available in the neutron-rich isotopes up to molybdenum~\cite{klaus}, there is essentially no $(\gamma,p)$ information 
in the neutron-deficient side of the nuclear chart, where $(\gamma,p)$ yields are expected to be relevant.   


\section*{Acknowledgements}

The author acknowledges fruitful physics discussions with  Cebo Ngwetsheni, Dorel Bucurescu,  Klaus Spohr, Balaram Dey  
and the late Denys Wilkinson.


\end{document}